\RequirePackage{lineno} 
\documentclass[twocolumn,prd,showpacs,preprintnumbers,amsmath,amssymb]{revtex4-2}
\usepackage{graphicx}
\usepackage{subcaption}
\usepackage{dcolumn}
\usepackage{bm}
\usepackage[usenames]{color}
\usepackage{multirow}
\usepackage{bigstrut}
\usepackage[utf8]{inputenc} 
\usepackage{ragged2e} 
\usepackage{url}
\usepackage[english]{babel}

\usepackage{float} 
\def\met{\mbox{\ensuremath{\, \slash\kern-.6emE_{\rm T}}}}

\begin{document}


\title{Real-time Accelerator Diagnostic Tools for the MAX IV Storage Rings}


\author{B. Meirose}
\email{bernhard.meirose@maxiv.lu.se}
\author{V. Abelin}%
\author{F. Bertilsson}%
\author{B.E. Bolling}%
\author{M. Brandin}%
\author{M. Holz}%
\author{R. H{\o}ier}%
\author{A. Johansson}
\author{P. Lilja}
\author{J. S. Lundquist}%
\author{S. Molloy}%
\author{F. Persson}%
\author{J. E. Petersson}%
\author{H. Serodio}%
\author{R. Sv{\"a}rd}%
\author{D. Winchester}%
\affiliation{MAX IV Laboratory, Lund University, \\ Fotongatan 2, 224 84 Lund}%

\date{\today}

\begin{abstract}
In this paper, beam diagnostic and monitoring tools developed by the MAX IV Operations Group are discussed. In particular, new beam position monitoring and accelerator tunes visualization software tools, as well as tools that directly influence the beam quality and stability are introduced. An availability and downtime monitoring application is also presented.

\end{abstract}


\maketitle
\section{Introduction} 
The MAX IV laboratory \cite{MAXIVsaga} is a synchrotron radiation facility in Lund, Sweden. MAX IV is the first Multi-Bend Achromat (MBA) Synchrotron Light Source in the world and provides scientists with the most brilliant X-rays for research. The laboratory was inaugurated in June 2016 and consists of two storage rings operated at 1.5 and 3~GeV providing spontaneous radiation of high brilliance over a broad spectral range. The 1.5~GeV ring has a circumference of 96~m and employs a double-bend achromat lattice to produce an emittance of 6~nm~rad. The 3~GeV storage ring on the other hand is aimed towards ultralow emittance to generate high brilliance hard X-rays. The design of the 3~GeV storage ring includes many novel technologies such as MBA lattice and a compact, fully-integrated magnet design. This results in a circumference of 528~m and an emittance as low as 0.2~nm~rad \cite{Pedro}. A linear accelerator (linac) works as a full-energy injector for the storage rings as well as a driver to a Short Pulse Facility (SPF). The prime sources for synchrotron radiation at the rings are optimized insertion devices (IDs), providing intense X-ray light for each of the MAX IV beamlines.

This article is organized as follows. Section \ref{operations} briefly summarizes the nominal operation conditions of the storage rings and of the SPF. In Section \ref{trends} a new software tool which displays the beam position's evolution over time
and the accelerator tunes' evolution over time using two-dimensional colormap spectrograms is presented. A tool that reduces the impact of changing orbit bumps for the MAX IV beamlines is presented in section \ref{wobbl}. Section \ref{morf} introduces an application for adjusting the master oscillator's radio-frequency of a storage ring that is nearly transparent for beamlines.  Section \ref{downtime} describes the MAX IV downtime web-application used for registering and monitoring the availability and downtime of the MAX IV accelerators with automatized plotting capabilities allowing for prompt statistical analysis of events. Conclusions are presented in Section \ref{conclusion}.

\section{MAX IV ACCELERATOR OPERATIONS} 
\label{operations}
The MAX IV storage rings \cite{Tavares2019} currently operate in a 30~minute top-up injection mode during user delivery. High bunch charge is delivered to the SPF in-between injections. Although both rings are filled in 30~minute intervals, different injection modes are used. In the 1.5~GeV storage ring injections are accomplished using a single dipole kicker magnet, which although efficient \cite{Leemann2012}, is not transparent to users leading to betatron oscillations of the stored beam with amplitudes of the order of several millimeters. In the 3~GeV storage ring a Multipole Injection Kicker (MIK) method is used, which delivers quasi-transparent injections with good capture efficiency \cite{tupgw063}. Under nominal conditions the 3~GeV storage ring operates with 166~filled RF buckets, out of a maximum (harmonic number) of 176, while the 1.5~GeV ring operates at 32~even-filled RF buckets.

Both storage rings employ passive harmonic (Landau) RF cavities \cite{Tavares2014}. The main  function  of  the  harmonic  cavities  is to generate a  voltage which counteracts  the longitudinal focusing at the bunch center of  the  main  RF voltage, lengthening  the  bunch,  decreasing  the charge density, thus damping instabilities and increasing the Touschek-dominated lifetime.

The 3~GeV and 1.5~GeV rings make use of a number of feedback systems to deliver stable and reliable beams. These include main cavity field amplitude and phase, main cavity frequency and Landau cavity field amplitude, all implemented in an FPGA-based low level RF-system \cite{LLRF}. The 3~GeV ring uses also feedback to damp Robinson mode oscillations \cite{MZD}. A digital Bunch-By-Bunch (BBB) feedback is used to damp coupled bunch modes and as a diagnostic tool \cite{BBB_R3, BBB_R1}.

A slow orbit feedback (SOFB) \cite{SOFB} loop is used in both storage rings to handle drifts. The main goal of this is to keep the electron orbit as constant as possible at all times. Using information from the BPMs, which measure the transverse position of the beam along the storage ring, dipole corrector magnets currents are adjusted so that the beam orbit is corrected towards offsets or a pre-defined golden orbit \cite{golden}. The SOFB loop updates at a 10~Hz rate. 

A feedback system that acts directly on the accelerator radio frequency master oscillator (MO RF) is used in order to keep the energy contribution from the storage ring's corrector magnets constant. In the 1.5~GeV ring a tune feedback is used in order to keep the betatron tunes close to a predefined optimal value. While the tune feedback system itself makes use of the ring's quadrupole magnets and pole-face strips that are installed in dipole magnets as actuators, the betatron tunes are themselves obtained by the ring's BBB system. The main feedback systems used at the MAX IV storage rings are shown in table \ref{Tab:R1_RF}. For completeness, a summary of key delivery parameters for the 1.5 and 3~GeV storage rings are presented in tables \ref{Tab:R1} and  \ref{Tab:R3}, respectively.



\begin{table}
\renewcommand{\tabcolsep}{1.2mm}
\caption{Feedback systems used in the MAX IV Storage Rings}
\label{Tab:R1_RF}
\begin{center}
\begin{tabular}{|c|cc|cc|}
\hline
MO RF feedback \\ 
\hline
Main cavity field amplitude and phase\\   
\hline
Main cavity frequency\\  
\hline
Landau cavity field amplitude\\ 
\hline
Robinson mode feedback (only 3~GeV ring) \\ 
\hline
Slow orbit feedback\\ 
\hline
Tune feedback (only 1.5~GeV ring) \\ 
\hline
\end{tabular}
\end{center}
\end{table}


\begin{table}
\renewcommand{\tabcolsep}{1.2mm}
\caption{Nominal operations conditions for the 1.5~GeV Storage Ring}
\label{Tab:R1}
\begin{center}
\begin{tabular}{|c|cc|cc|}
\hline
\textbf{Parameter} & \textbf{Value} &\\
\hline
Stored current & 500~mA &\\ 
\hline
Top-up mode & 30~min via dipole kicker &\\ 
\hline
Filling pattern & Uniform (32~buckets) &\\ 
\hline
Landau cavities conditions & 2~cavities tuned in &\\ 
\hline
BBB feedback & ON (vertical and horizontal) &\\ 
\hline
Slow orbit feedback & ON &\\ 
\hline
\end{tabular}
\end{center}
\end{table}

\begin{table}
\renewcommand{\tabcolsep}{1.2mm}
\caption{Nominal operations conditions for the 3~GeV Storage Ring}
\label{Tab:R3}
\begin{center}
\begin{tabular}{|c|cc|cc|}
\hline
\textbf{Parameter} & \textbf{Value} &\\
\hline
Stored current & 250~mA &\\ 
\hline
Top-up mode & 30~min via MIK &\\ 
\hline
Filling pattern & 166/176~filled buckets &\\ 
\hline
Landau cavities conditions & 3~cavities tuned in &\\ 
\hline
BBB feedback & ON (vertical) &\\ 
\hline
Slow orbit feedback & ON &\\ 
\hline
\end{tabular}
\end{center}
\end{table}

The current nominal operation conditions for the SPF~\cite{Thorin2014} aim at delivering 100~pC charge per bunch at a repetition rate of 2~Hz. To achieve this, electrons are extracted from a copper photo-cathode using a laser pulse, then accelerated to 3~GeV. The bunches are compressed in two stages, at 233~MeV and at the full 3~GeV, to about 100~fs. Then the beam goes through the two undulators of the SPF, generating equally short x-ray pulses. As in the rings, there is a feedback on the trajectory, keeping the beam at the center of the optics.

\begin{figure}
\includegraphics[scale=0.18]{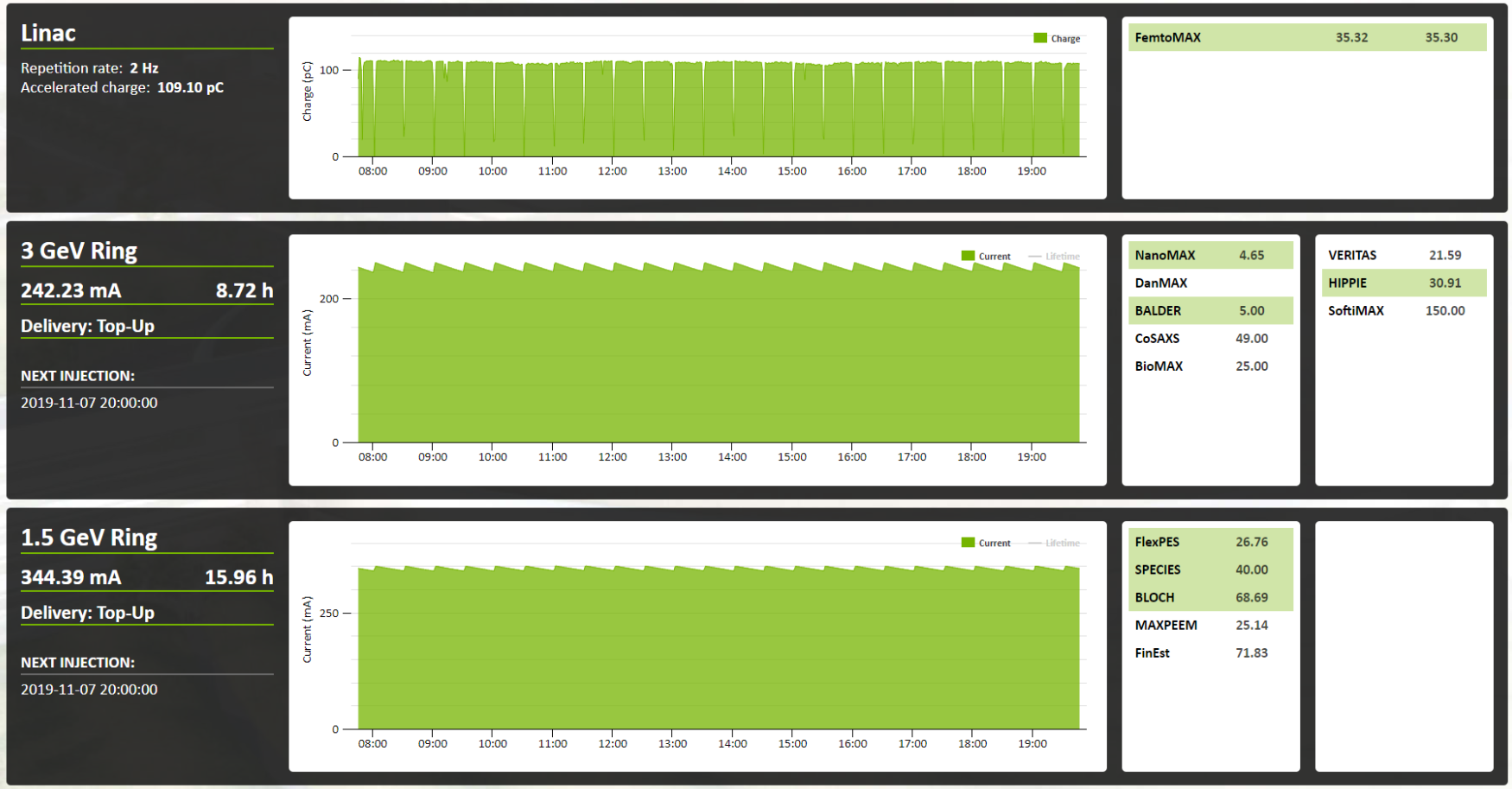}
\caption{\label{fig:status} MAX IV status page in November 7, 2019.}
\end{figure}

Figure ~\ref{fig:status} shows the facility status in November 2019. The status page shows the current stored in the two storage rings and the charge delivered to the SPF at present time and the evolution of the stored current during the previous 12~hours. It also shows the positions of the undulator gaps and the status of the beam shutters at all beamlines. A closed gap means that synchrotron light is produced by a beamline's insertion device whilst open shutters, indicated by highlighting the beamline name in green, means that the beamline is open to take the light produced by its insertion device. 
As can be seen from the plot, the linac performed excellently as injector providing beam current top-up every 30~minutes, as well as  high bunch charge (100~pC) to the SPF.

\par

\section{Beam Position Monitor Time Evolution} 
\label{trends}
\begin{figure*}[!tbh]
    \centering
    \captionsetup{width=.79\linewidth}
    \includegraphics*[width=0.79\textwidth]{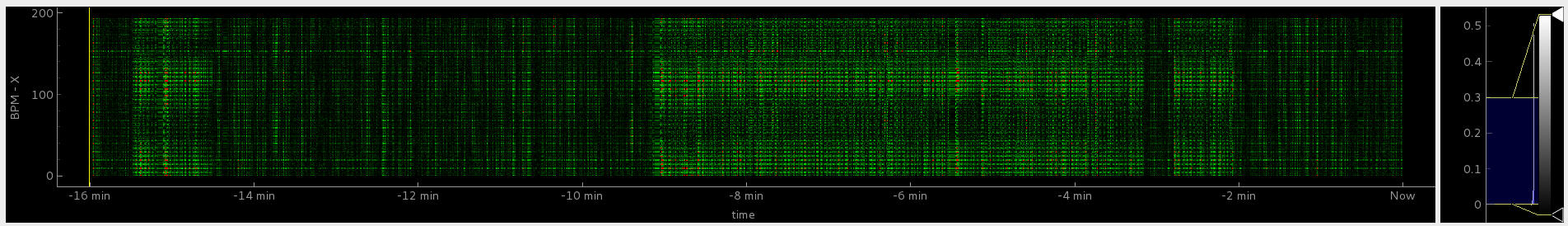} 

    \caption{BPM trends in the horizontal plane for the 3~GeV storage ring during stable-beam delivery.}
    \label{fig:BPM_trend_normal}
\end{figure*}

\begin{figure*}[!tbh]
    \centering
    \captionsetup{width=.79\linewidth}
    \includegraphics*[width=0.79\textwidth]{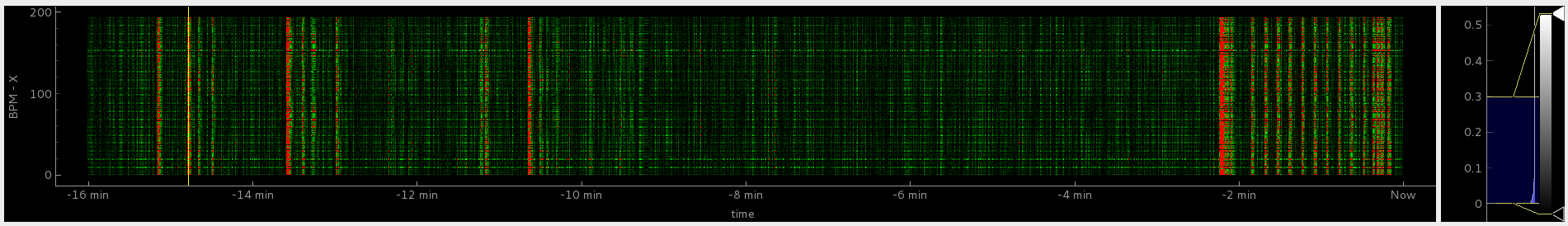} 

    \caption{BPM trends in the horizontal plane for the 3~GeV storage ring. The vertical lines at before -10~min and around and beyond -2~min are due undulator gap movements from a beamline.}
    \label{fig:BPM_beamline}
\end{figure*}

\begin{figure*}[!tbh]
    \centering
    \captionsetup{width=.79\linewidth}
    \includegraphics*[width=0.79\textwidth]{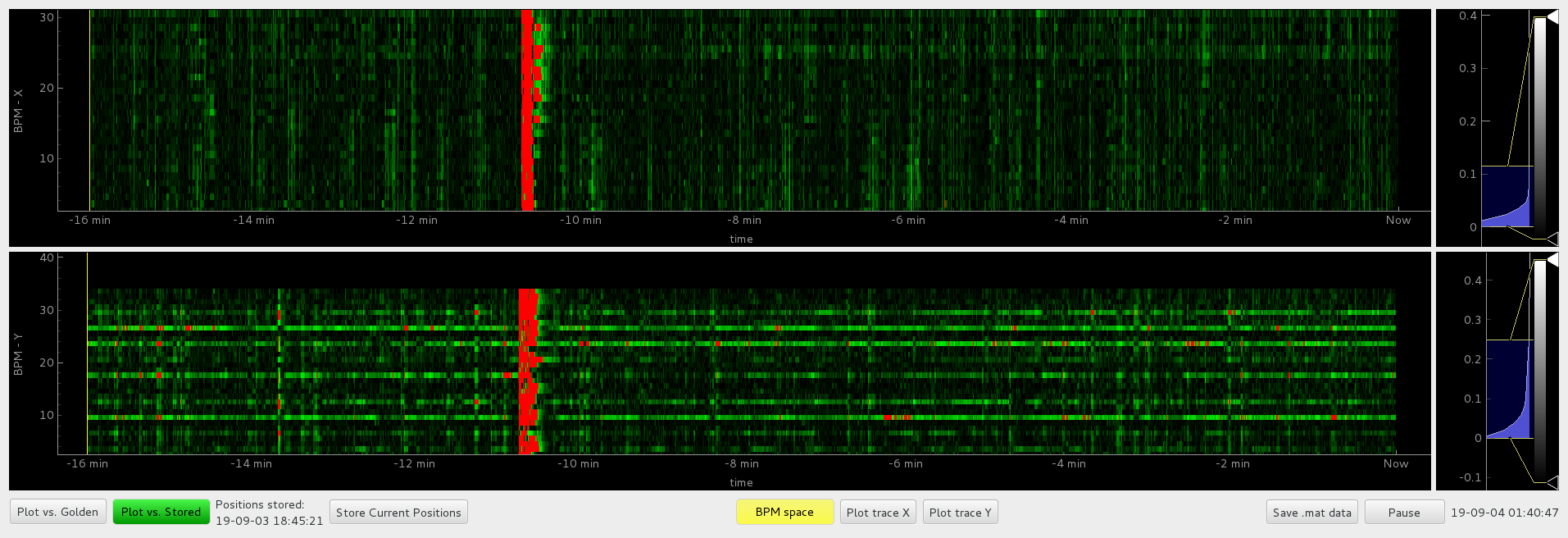} 

    \caption{BPM trends in the horizontal (top) and vertical (bottom) planes for the 1.5~GeV storage ring. The thick vertical lines seen in both planes are due to dipole kicker top-up injections.}
    \label{fig:BPM_R1}
\end{figure*}

The Beam Position Monitor Time Evolution (BPM trends) is a tool developed by the operations group that shows the storage rings' beam positions over time (a real-time ``sliding plot'' during operations) as measured by all the rings' BPMs. The vertical scale on the right (in $\mu$m) controls whether there is any deviation within the lower and upper limits set by the scale. Figure~\ref{fig:BPM_trend_normal} shows a typical example of stable delivery, where all BPM readings show the beam is kept within the required limits (within 0~$\mu$m to $\pm$0.3~$\mu$m, in this case) and are therefore shown in green. When the deviation is above the upper limit (0.3~$\mu$m) the points are shown in red. An illustration of this fact is demonstrated in Figure~\ref{fig:BPM_beamline}, where one can observe that small changes to one of the beamline's gaps in the 3~GeV storage ring can cause visible disturbances to the beam. In Figure~\ref{fig:BPM_R1} it is noticeable that during injections a large vertical line can be observed in the 1.5~GeV ring BPM trends, due to the significant disturbances caused by the dipole kicker that it is used in the smaller ring. A similar line is not observed in Figure~\ref{fig:BPM_trend_normal} as the 3~GeV ring makes use of the MIK.

The tool allows also the possibility of switching mode between "BPM-space" and "corrector-space", the former being the positions measured by the BPMs, and the latter being the BPM positions transformed through the orbit response matrix (ORM) to the angular kicks at the corrector magnets that would result in the observed orbit distortions. The ORM is normally used for the feedback system to calculate corrector magnet strengths, and by the same logic, it is possible to use the ORM to locate where kicks on the beam are originating from.  

\begin{figure*}
    \centering
    \captionsetup{width=.7\linewidth}
    \includegraphics*[width=0.7\textwidth]{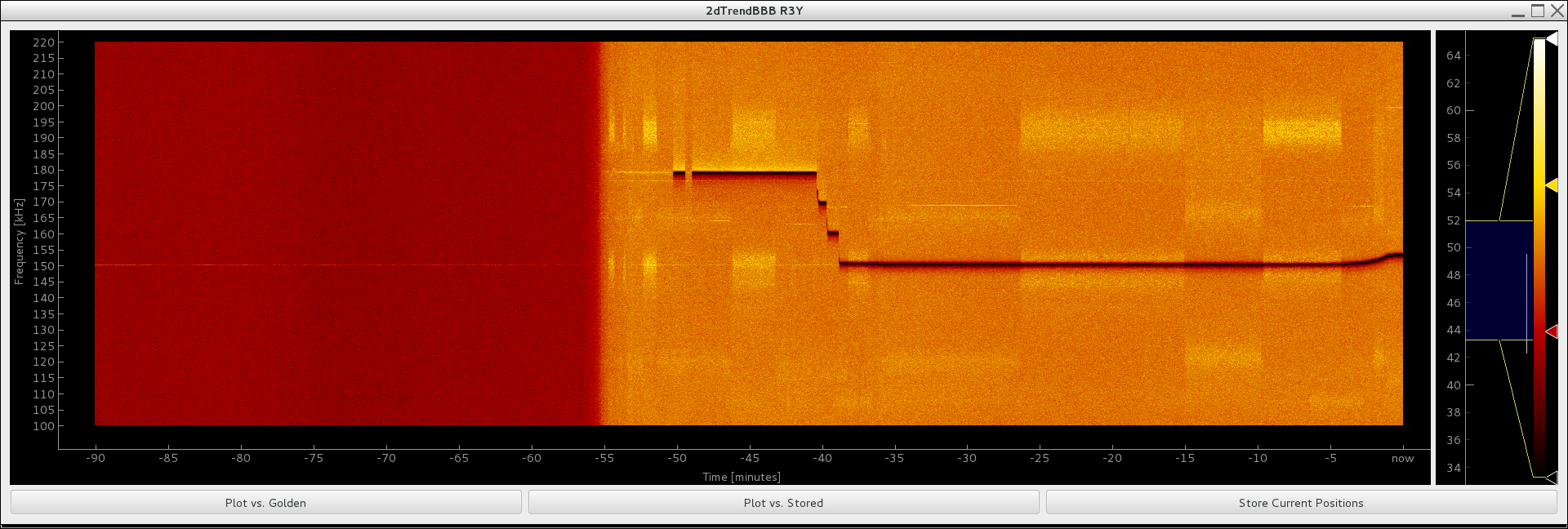} 

    \caption{Figure showing (starting approximately at -40~min) a manual increase of the Master Oscillation (MO) RF of the 3~GeV storage ring in steps of 5~Hz in order to change the vertical betatron tune.}
    \label{fig:Frequency_plot}
\end{figure*}

\begin{figure*}
    \centering
    \captionsetup{width=.79\linewidth}
    \includegraphics*[width=0.79\textwidth]{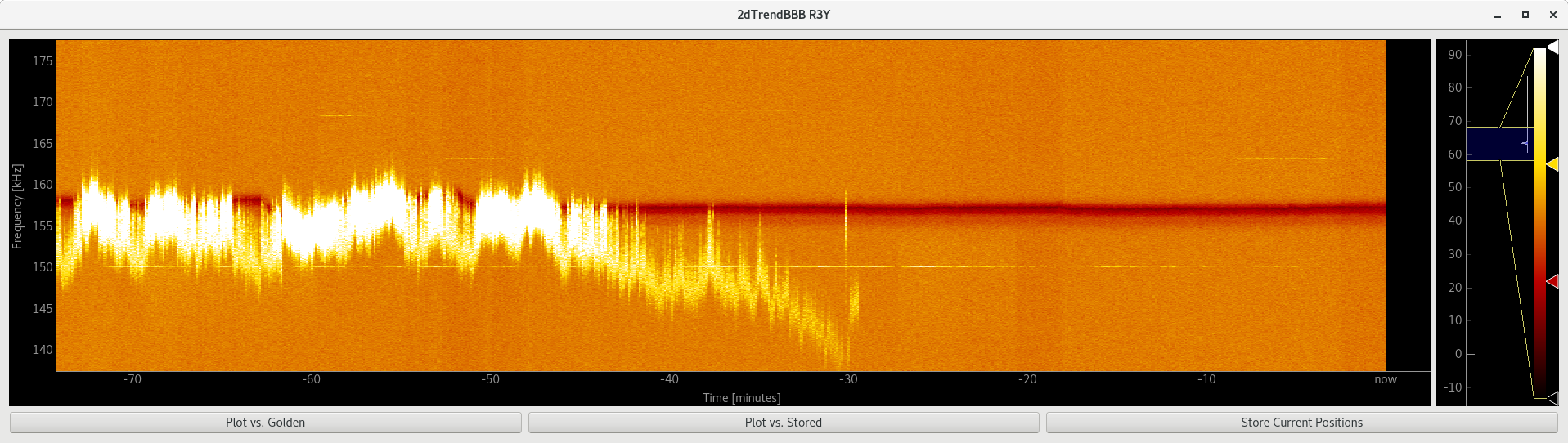} 

    \caption{Accelerator tunes visualization in the 3~GeV storage ring. Between -80 and -40~min, a disturbance can be seen around the tune frequency. At -40~min a faulty amplifier connected to a tune measuring system was turned off. This vertical disturbance caused an increase in vertical emittance, which decreased to nominal values as soon as the disturbance moved away from the tune frequency.}
    \label{fig:Frequency_plot_unstable}
\end{figure*}

Two-dimensional colormap spectrograms are also useful to monitor the synchrotron and betatron frequencies variations measured by the BBB system \cite{BBB_R3, BBB_R1} and, therefore, a useful tool to monitor the accelerator's tunes. At times when there are disturbances of the same order as the frequency of the tunes, the spectrograms convey a clearer image than just observing the tune value. Figure \ref{fig:Frequency_plot} illustrates a typical use case. At approximately -40~min, the MO RF of the 3~GeV storage ring is manually increased in an attempt to change the vertical tune closer to its delivery set value (0.265). The MO RF changes are done in 5~Hz steps, which is visible in the staircase decrease of the vertical betatron frequency. At -5~minutes one of the beamlines closes its undulator gap, causing the tune to shift, which is clearly visible at right end of the plot. 
An example of the use of this tool to diagnose faulty equipment is given in  Figure \ref{fig:Frequency_plot_unstable}. Between -80 and -40~min, a disturbance can be seen around the tune frequency. At -40~min a faulty amplifier connected to 
an alternative tune measurement system was turned off. This vertical disturbance caused an increase in vertical emittance, which decreased to nominal values as soon as the disturbance moved away from the tune frequency.

\begin{figure*}
    \centering
    \captionsetup{width=.7\linewidth}
    \includegraphics*[width=0.7\textwidth]{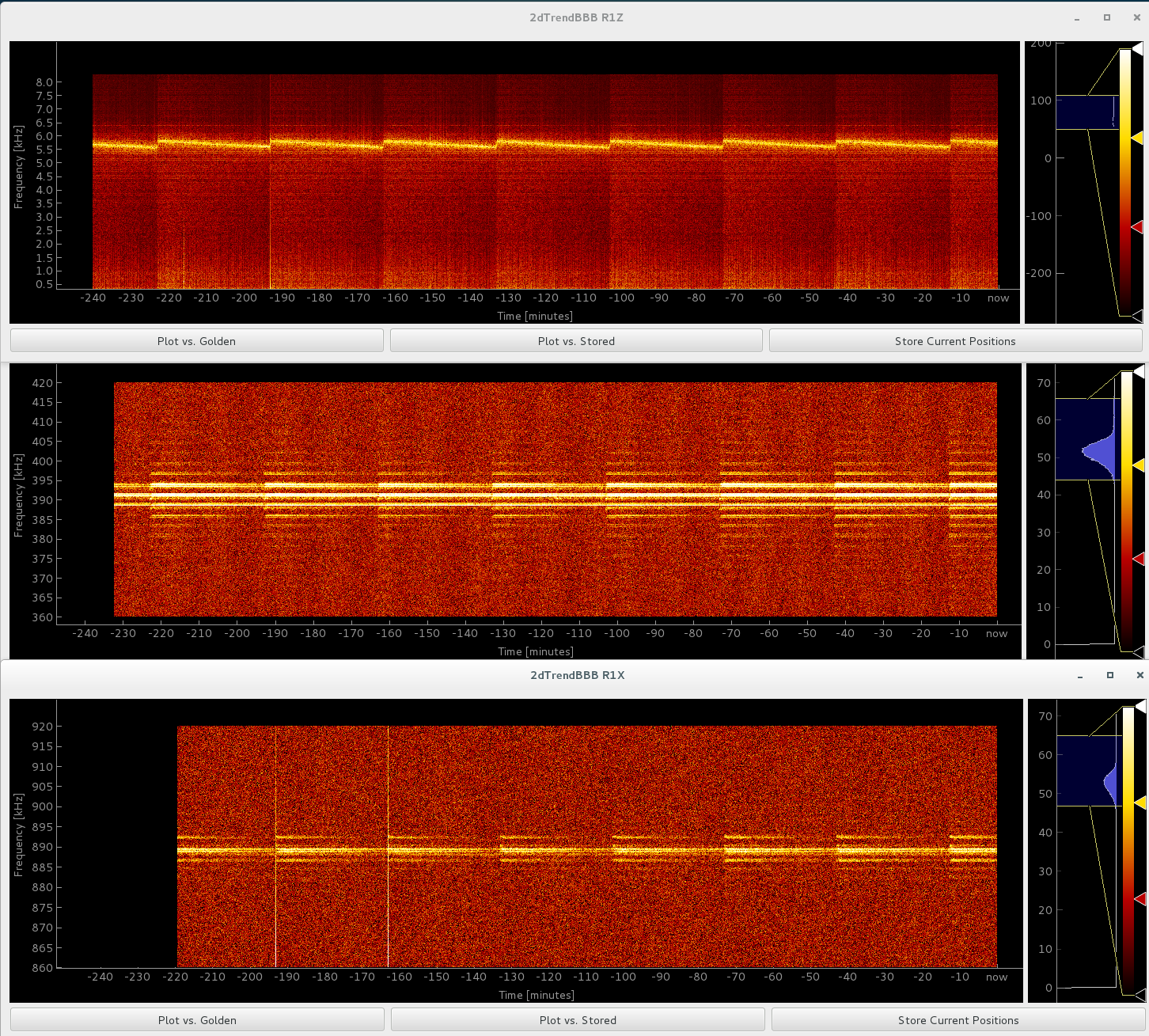} 

    \caption{The longitudinal and transverse frequencies for the 1.5~GeV storage ring. In the longitudinal plane (synchrotron frequency) one can see the effects of the top-up injections. In the vertical plane (middle plot) one can see the signal is strongest around 350~kHz, but lines for higher and lower frequencies, albeit weaker, can be seen for higher and lower frequencies. A similar, but less pronounced effect can be observed in the horizontal plane (bottom plot).}
    \label{fig:Frequency_plot_AJ_R1}
\end{figure*}

The tool also provides a good way to visualize whether the beam is behaving as expected in all planes of space.
Figure \ref{fig:Frequency_plot_AJ_R1} shows the longitudinal and transverse frequencies for the 1.5~GeV storage ring. In the longitudinal plane (synchrotron frequency) one can see the effects of the top-up injections every 30~minutes. In the vertical plane (middle plot) one can see the signal is strongest around 350~kHz, but lines for higher and lower frequencies, albeit weaker, can be seen, which is consistent with the synchrotron sidebands of the betatron frequency measured by the BBB system. A similar, but less pronounced effect can be observed in the horizontal plane (bottom plot). Such plots can be constantly monitored during beam delivery, and should significant deviations on the shown patterns be observed in any of the planes, actions can be taken to improve beam quality and prevent beam losses.


\section{WOBBL} 
\label{wobbl}
Changes to the orbit of the beam through the long straight sections of the storage rings where the insertion devices are situated, are sometimes requested by beamline users at MAX IV during normal operation.

The implementation of such orbit bumps is done by changing the reference orbit to which the SOFB corrects the beam. Manual implementation of a beamline bump results in an abrupt disturbance of the stored beam, which is not transparent to the users at the beamlines.

In order to accommodate beamline bump change requests without disturbing the measurements performed at any of the other beamlines, a Widget for setting the Orbit Bumps for BeamLines (WOBBL) has been developed within the operations group.

WOBBL implements changes to the reference positions for the SOFB in a series of sub-micrometer steps at a repetition rate which is lower than that of the feedback system. This allows the orbit to be corrected in-between each step change. The result is a smooth transition from the initial orbit to the desired final orbit.

\begin{figure*}[!htb]
\includegraphics*[width=.79\textwidth]{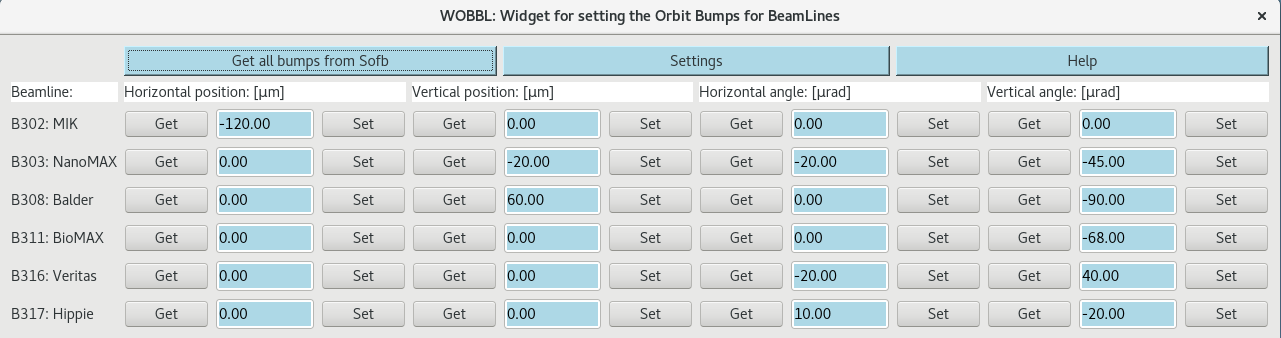} 
\caption{WOBBL interface containing the beamlines which are in normal operation for the 3~GeV ring and also the MIK. The widget has input fields for setting each of the four types of bump changes, for each beamline.}
\label{fig:wobbl}
\end{figure*}

Figure \ref{fig:wobbl} shows the WOBBL interface which is used to implement the beamline orbit changes non-disruptively. The WOBBL interface shows the beamlines which are in normal operation for the 3~GeV ring and also the MIK. The widget has input fields for setting each of the four types of bump changes, for each beamline. Via the widget it is also possible to get a read-back of the positions and angles through the insertion devices, as picked up by BPMs in the ring. Both the step size and repetition rate are available as input settings.
 
In order to determine the effectiveness of the WOBBL tool in a controlled manner, studies were performed for the NanoMAX \cite{NanoMAX} beamline, one of the most instability-sensitive beamline at the MAX IV laboratory. During the measurements, the bumps for another beamline (BioMAX \cite{BioMAX}) were changed both directly and via the WOBBL tool and compared. 

\begin{figure*}[!htb]
\includegraphics*[width=.79\textwidth]{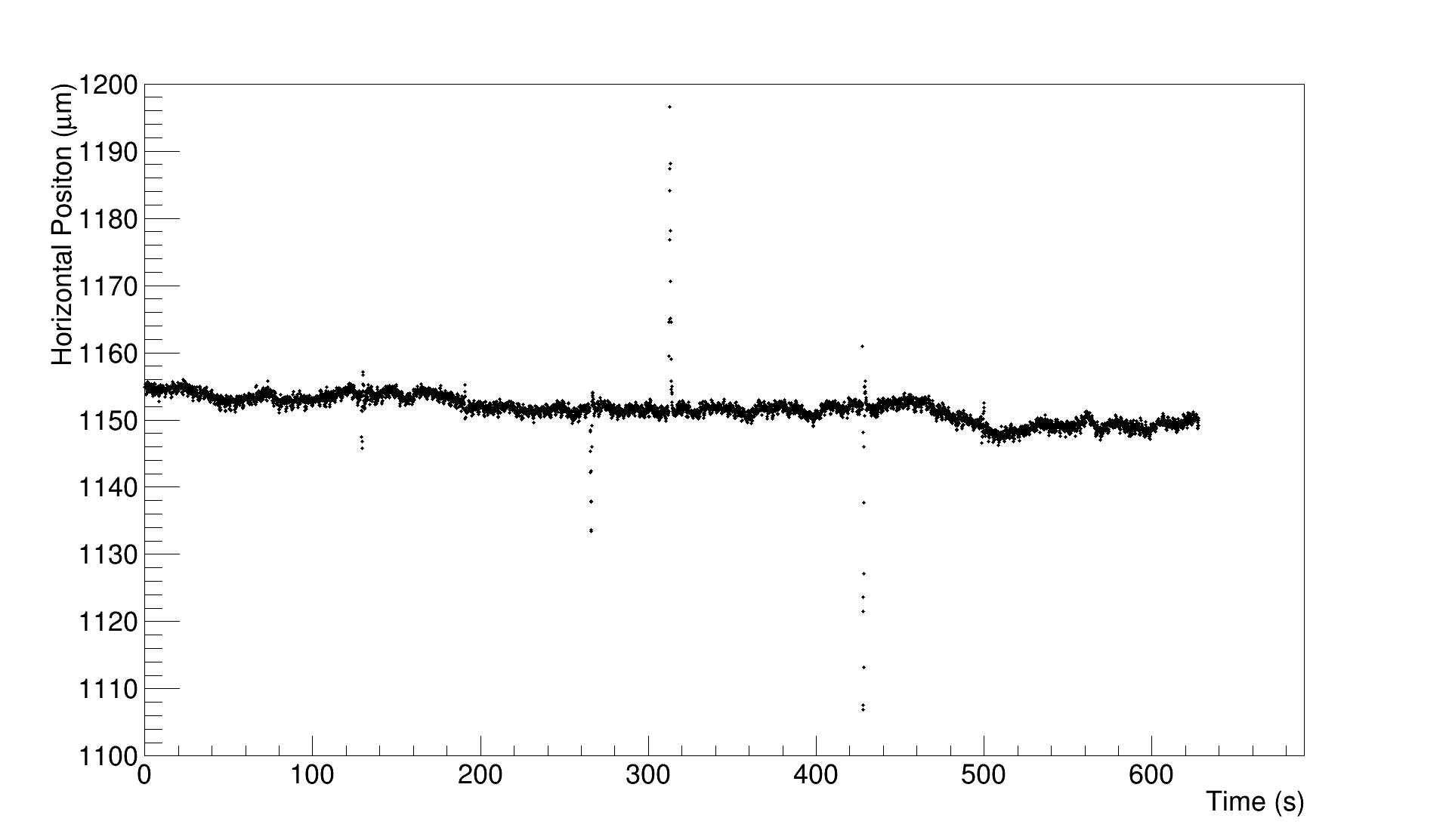}
\caption{Horizontal position observed by NanoMAX. The points where large perturbations are observed, correspond to direct bump changes performed in the same plane for the BioMAX beamline. See main text for details.}
\label{fig:Horizontal_Position_wobblMat}
\end{figure*}

\begin{figure*}[!htb]
\includegraphics*[width=.79\textwidth]{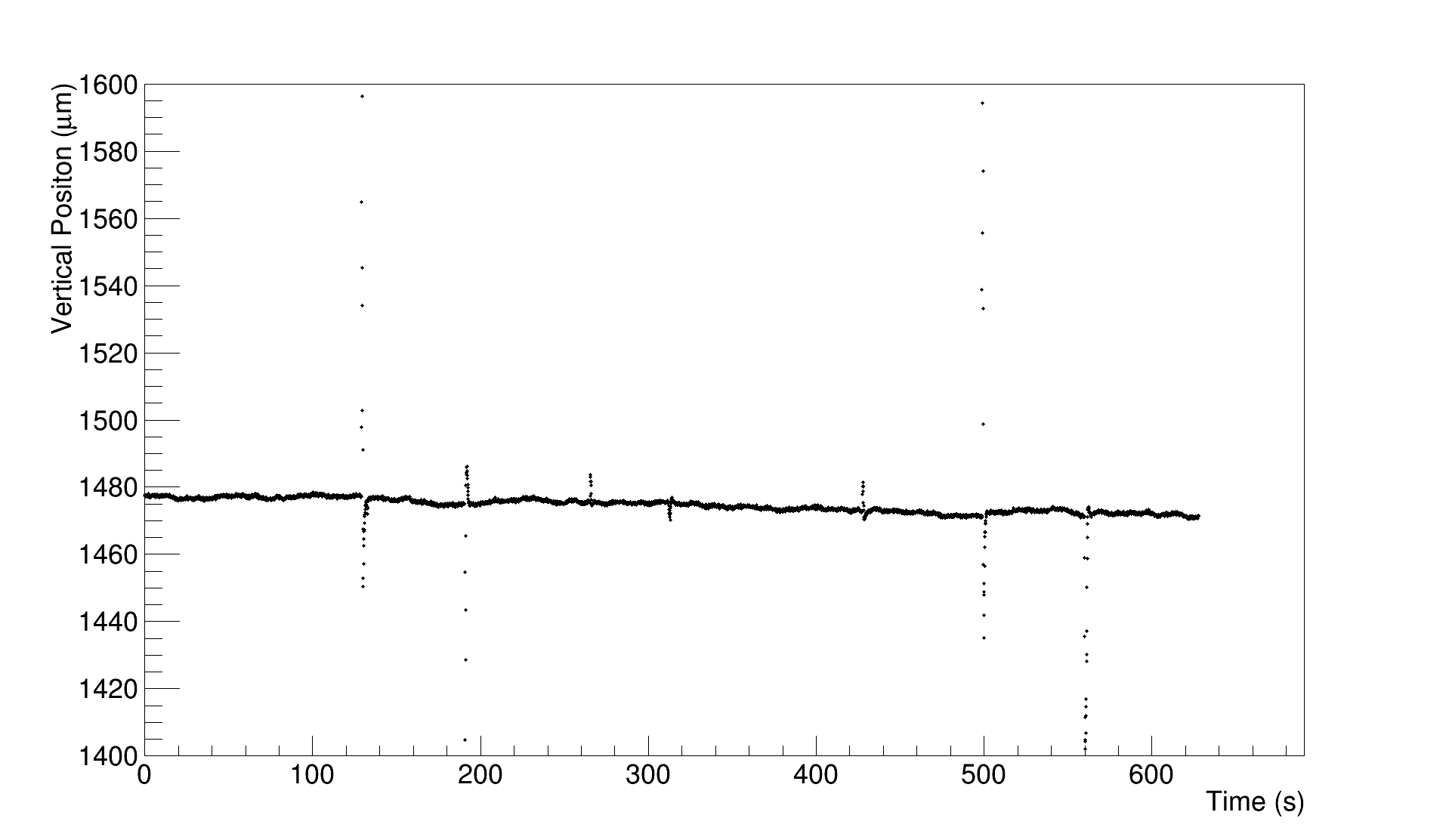}
\caption{Vertical position observed by NanoMAX. The points where large perturbations are observed, correspond to direct bump changes performed in the same plane for the BioMAX beamline. See main text for details.}
\label{fig:Vertical_Position_wobblMat}
\end{figure*}

Figure~\ref{fig:Horizontal_Position_wobblMat} shows the horizontal position observed by NanoMAX. At $t$ = 129~s, the vertical angle of the BioMAX beamline is directly increased by 68~$\mu$m. While disturbunces in the horizontal plane are relatively small, a significant positive vertical displacement can be seen in Figure \ref{fig:Vertical_Position_wobblMat}. At $t$ = 190~s, BioMAX'S horizontal angle is now decreased by 68~$\mu$m and, consistently, a negative displacement is seen by the beam delivered to NanoMAX.  At $t$ = 265 the horizontal angle is increased, and subsequently decreased (at $t$ = 312~s), but since the changes are performed in a perpendicular plane, the disturbances are much less pronounced. Other horizontal ($t$ = 427~s) and vertical (at $t$ = 498~s and $t$ = 559~s) bump changes are done directly at BioMAX and again the effects are consistently visible. In Fig. \ref{fig:Vertical_Position_wobblMat} it is noticeable that all the direct vertical changes mentioned for the BioMAX's bumps are unsurprisingly more visible for NanoMAX's measured vertical position, while the horizontal changes appear as small perturbations. 


\begin{figure*}[!htb]
\includegraphics*[width=.79\textwidth]{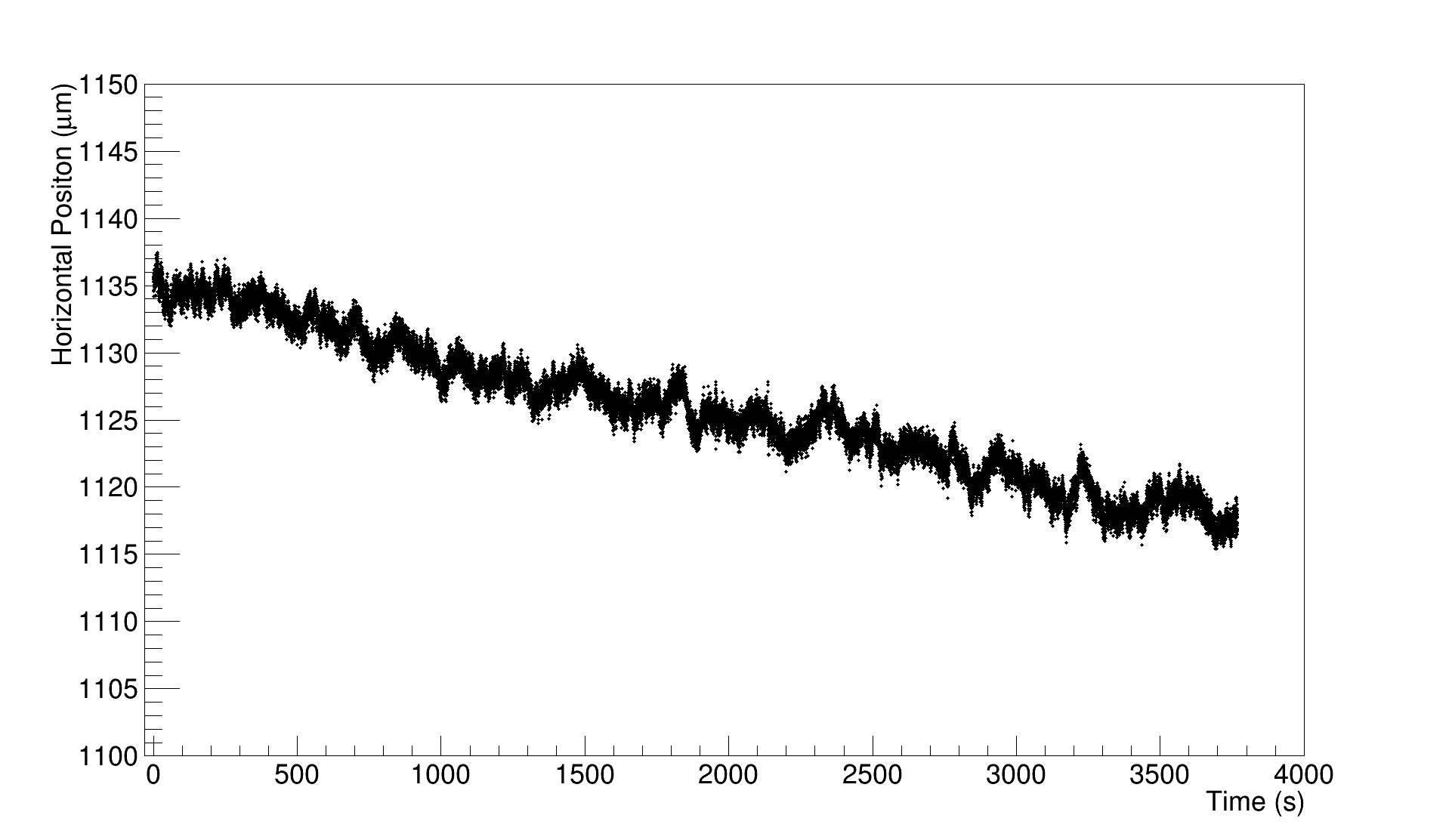}
\caption{Horizontal position measured by NanoMAX during a period of approximately one hour, during which several changes are done at BioMAX's bumps now using the WOBBL tool. No significant perturbations can be observed due to the applied changes. See main text for details.}
\label{fig:Horizontal_Position_wobbl}
\end{figure*}

Figure \ref{fig:Horizontal_Position_wobbl} shows for comparison the horizontal position measured by NanoMAX during a period of approximately one hour, during which several changes (also of 68~$\mu$m for consistency with the direct changes) are done at BioMAX's bumps now using the WOBBL tool.  Vertical changes at BioMAX bumps are done at $t$ = 205~s, $t$ = 325~s, $t$ = 505~s, $t$ = 745~s, $t$ = 2365~s, $t$ = 2665~s and $t$ = 3085~s. Horizontal bump changes for BioMAX are done at $t$ = 925~s, $t$ = 1285~s and $t$ = 1945~s. No large disturbances in NanoMAX's beam position can be observed during those changes and a similar plot for NanoMAX's vertical position is not shown for concision reasons, since the only remarkable thing about it is the absence of any disturbances, just like in the horizontal position. It is worth mentioning that during the nearly one hour of data taking, NanoMAX's observed beam position drifts downwards by nearly 20~$\mu$m, but such a deviation is within the normal limits of perturbations observed at the NanoMAX beamline, considering the period of observation, and cannot, therefore, unambiguously be attributed to the WOBBL tool or disentangled from other possible causes, such as for example, temperature changes.




\section{SOFT MO RF SWEEPER} 
\label{morf}
The soft MO RF sweeper is a tool for changing the MO RF of a storage ring in combination with an orbit feedback system. With this tool, changes in the MO RF can be carried out without significantly affecting beam instability-sensitive beamlines as the adjustments made in each step are carefully set up such that the beam instabilities arising from the MO RF adjustments are as small as the regular beam oscillations. As with the WOBBL, each adjustment also gives time for the orbit feedback system to adjust the beam orbit after adjusting the MO RF. This is accomplished by fulfilling two conditions: a) at least one second has to elapse and b) at least five iterations of beam trajectory corrections has to be carried out before the next MO RF adjustment is done.

At the two MAX IV storage rings, it was found that MO RF adjustments of 0.05~Hz results are practically undetectable by beam instability-sensitive beamlines. 
The simple graphical user interface (GUI) of this tool shows which storage ring is selected, the current MO RF and the desired change of the MO RF. After activating the sweep, the MO RF is moved in steps of 0.05~Hz until the defined change has been accomplished or if stopped by the user.

\begin{figure*}[!tbh]
\centering
\captionsetup{width=.6\linewidth}
\includegraphics*[width=0.6\textwidth]{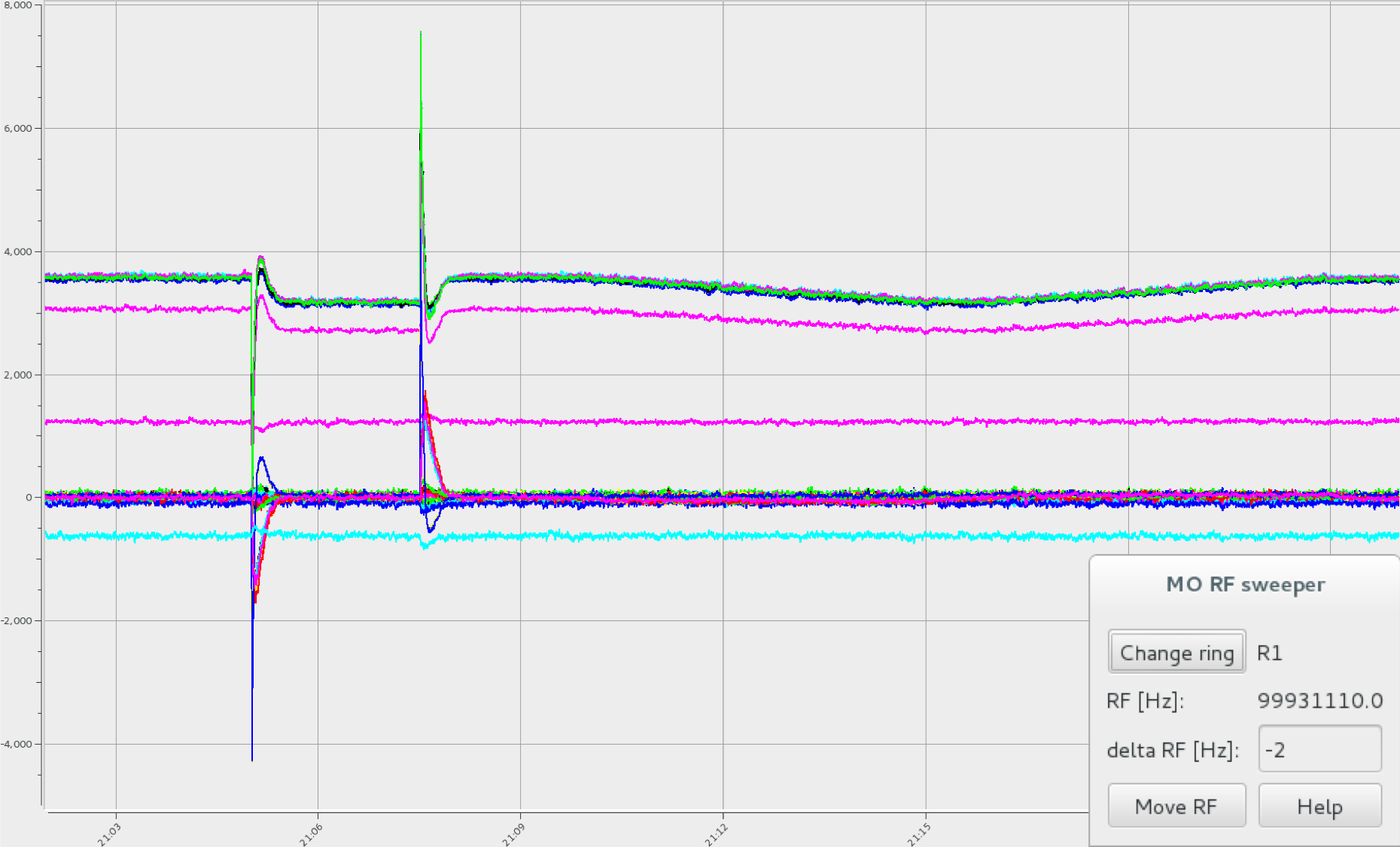}
\caption{The soft Master Oscillator RF in use. At 21.05 the MO RF is increased by 5~Hz directly (without the sweeper) and it is again decreased by 5~Hz directly at 21:07. A clear disturbance in the beam is measured by the BPMs.  The tool's GUI can be seen in the bottom-right corner.}
\label{fig:SWEEPER}
\end{figure*}

\begin{figure*}[!tbh]
\centering
\captionsetup{width=.6\linewidth}
\includegraphics*[width=0.6\textwidth]{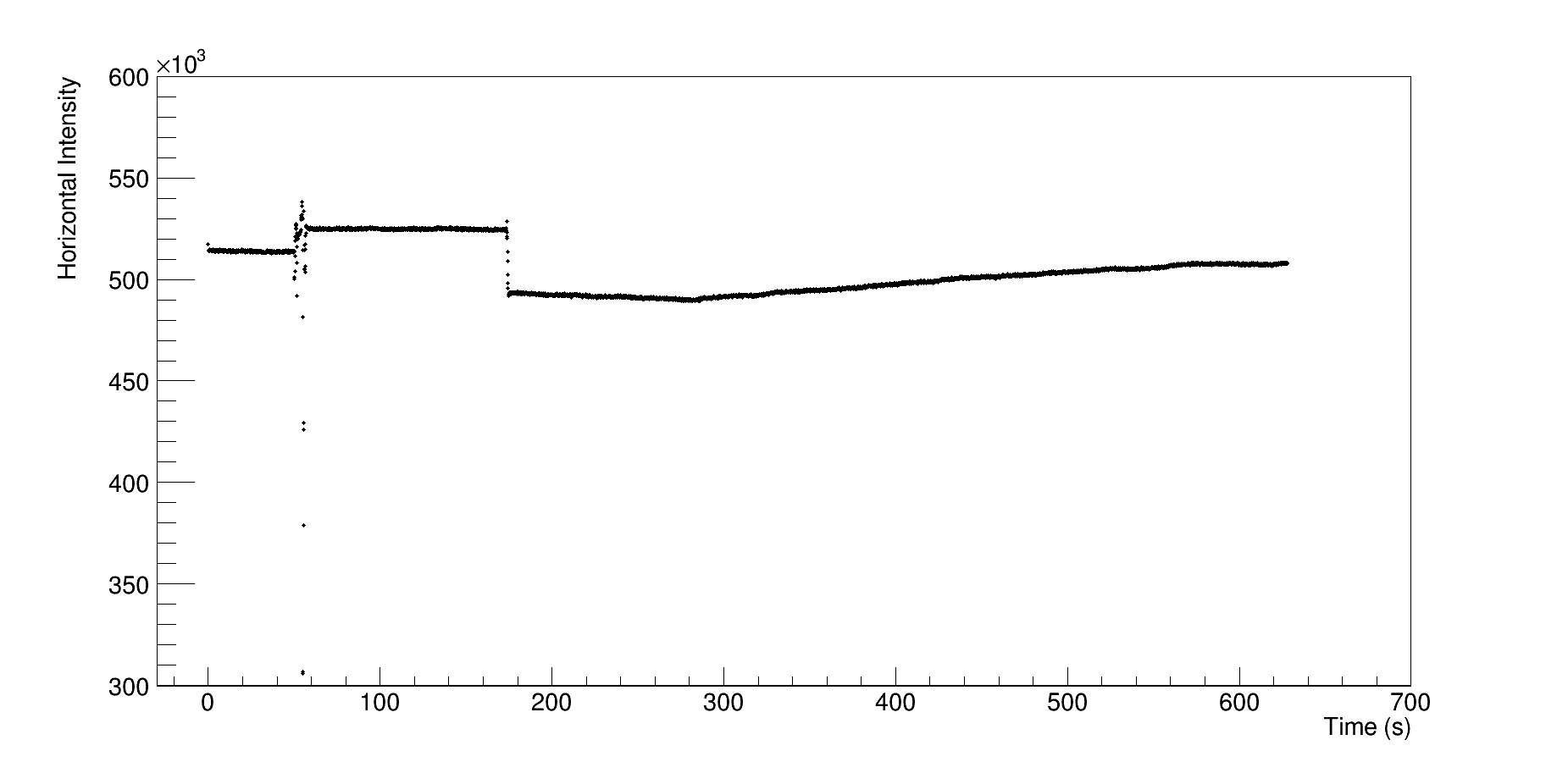}
\caption{NanoMAX's horizontal beam intensity. At t = 50~s the MO RF is increased directly by 5~Hz. The beam intensity decreases abruptly and significantly before rising again, as a result of MO RF increase. At t = 174~s, the MO RF is decreased by 10~Hz directly and once more a large disturbance in the delivered beam for NanoMAX is observed. At 285~s the MO RF starts to be increased towards a total 5~Hz change using the sweeper. The effect is observable as a subtle drop in the horizontal intensity at the recorded change time, which starts increasing again during the approximate 5~minutes it takes the MO RF to reach its new desired value.}
\label{fig:NanoMAX_HI}
\end{figure*}

\begin{figure*}[!tbh]
\centering
\captionsetup{width=.6\linewidth}
\includegraphics*[width=0.6\textwidth]{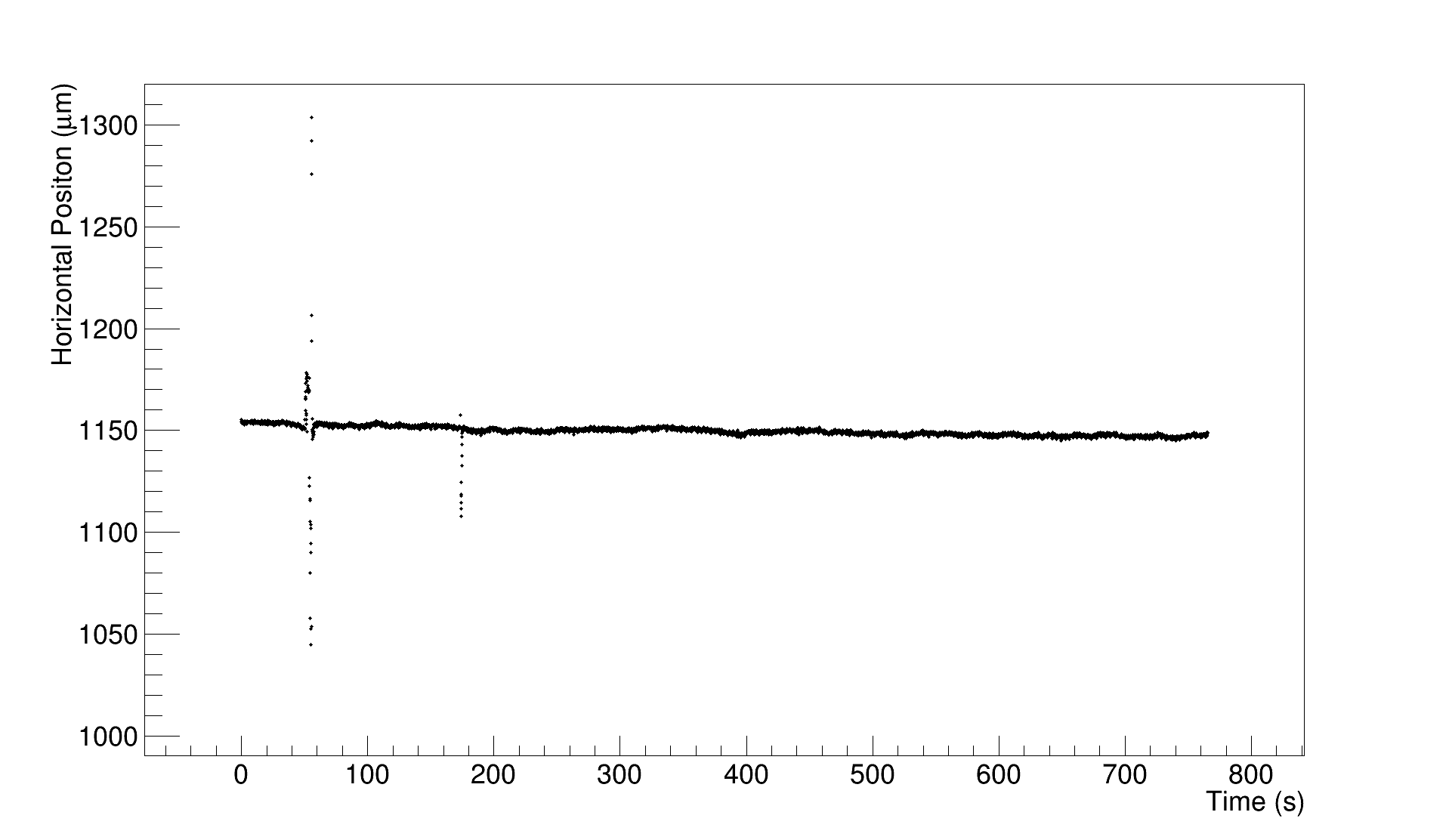}
\caption{NanoMAX's horizontal beam position. At t = 50~s the MO RF is increased directly by 5~Hz. The horizontal position oscillates as much as $\sim$ 150~$\mu$m  as a result of MO RF increase. At t = 174~s, the MO RF is decreased by 10~Hz directly and once more a large disturbance in the delivered beam for NanoMAX is observed. At 285~s the MO RF starts to be increased towards a total 5~Hz change using the sweeper, which takes approximately 5~minutes, but the change is not noticeable in the plot.}
\label{fig:NanoMAX_HP}
\end{figure*}

In Figure~\ref{fig:SWEEPER}, the difference between applying a change of 5~Hz directly and via the soft MO RF sweeper is visible from the 1.5~GeV storage ring beam data. At time instants when the MO RF is directly increased by 5~Hz (21:05) and decreased by 5~Hz (at 21:07) a large deviation in beam position is measured by the 1.5~GeV storage ring BPMs', both in the vertical and horizontal planes. At 21:10 the same 5~Hz increase is applied to the MO RF, followed by a 5~Hz decrease at 21:21, both via the sweeper - disturbances to the beam orbit are then barely seen. This preservation of stability that is accomplished with the MO RF sweeper is crucial during delivery for the experiments measurements to be carried out without disturbances. As with the WOBBL tool, the effect can be observed for NanoMAX, as shown in Fig.~\ref{fig:NanoMAX_HI}. At t~=~50~s the MO RF is increased directly by 5~Hz which can be observed both at NanoMAX's measured beam horizontal intensity,  as well as in its measured beam horizontal position. The beam intensity decreases abruptly and significantly before rising again, as a result of MO RF increase, while the horizontal position oscillates as much as $\sim$ 150~$\mu$m (a similar effect is observed in the vertical plane, but is not shown). At t = 174~s, the MO RF is decreased by 10~Hz directly and once more a large disturbance in the delivered beam for NanoMAX is observed for both quantities. At 285~s the MO RF starts to be increased towards a total 5~Hz change using the sweeper. The effect is observable as a subtle drop in the horizontal intensity at the recorded change time, which starts increasing again during the approximate 5~minutes it takes the MO RF to reach its new desired value. The change is not noticeable in the horizontal position plot in Figure~\ref{fig:NanoMAX_HP}, which is a strong indicator of the sweeper's usefulness during beam delivery.

\section{Availability and downtime monitoring} 
\label{downtime}
In order to improve the gathering of statistics concerning downtime duration and facilitate reliability studies, a ``downtime web-application'' was developed \cite{Rasmus}. This application has since its deployment proven itself useful for analyzing the different downtime events and for finding ways to prevent similar incidents in order to further reduce downtime. The application allows operators to easily edit and add accelerator operations schedules and report any unscheduled downtime to a database. For each downtime event, a machine (the Linac (I), the 3~GeV (R3) or the 1.5~GeV (R1) storage rings) is indicated together with an event type code, the date and time of the day, the downtime's duration and the name of the reporting operator together with a description of what happened. The application then uses the schedule and the downtime events to calculate Mean Time Between Failures (MTBF), Mean Time To Failure (MTTF) and Mean Time to Repair (MTTR) on timescales controlled by the user. It also allows easy visualization of downtime duration by machine (R1, R3, I) or event label, which can be beamlines, beam instability, controls, diagnostics, human error, high level software, injector, insertion devices, infrastructure, laser (photo-cathode gun), magnets, machine protection system, network, orbit interlock, programmable logic controller, personnel safety system, radio frequency, vacuum, water, WatchDog (software), or others, if the cause is unknown or no existing label is well suited.
\par
\begin{figure*}[!htb]
\includegraphics*[width=0.8\textwidth]{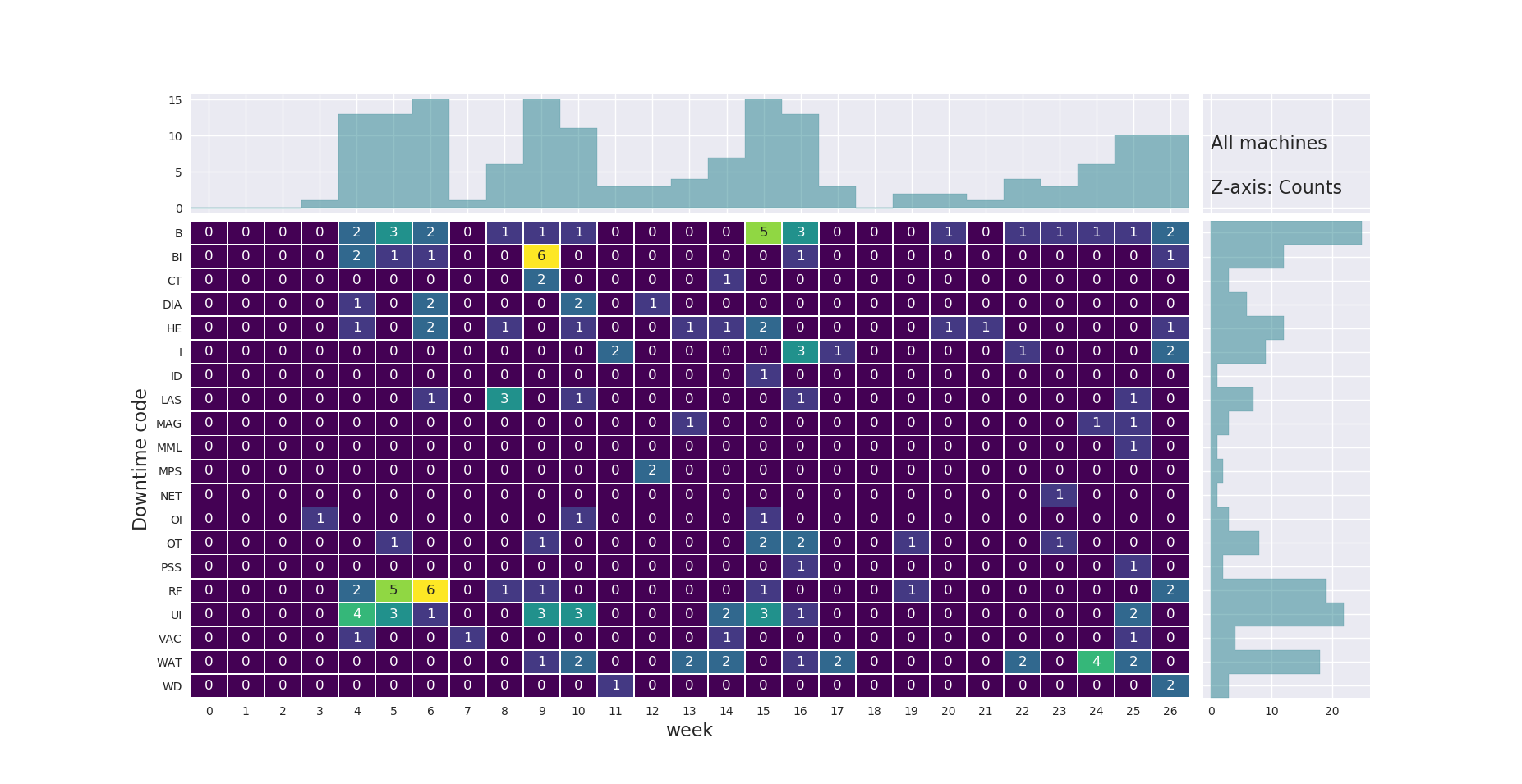} 
\captionsetup{width=.68\linewidth}
\caption{Heatmap of downtime at MAXIV as a function of event code and week number for all three machines (I, R3, R1).}
\label{fig:downtimeCodeWeek}
\end{figure*}
\begin{figure*}[!htb]
\includegraphics*[width=0.8\textwidth]{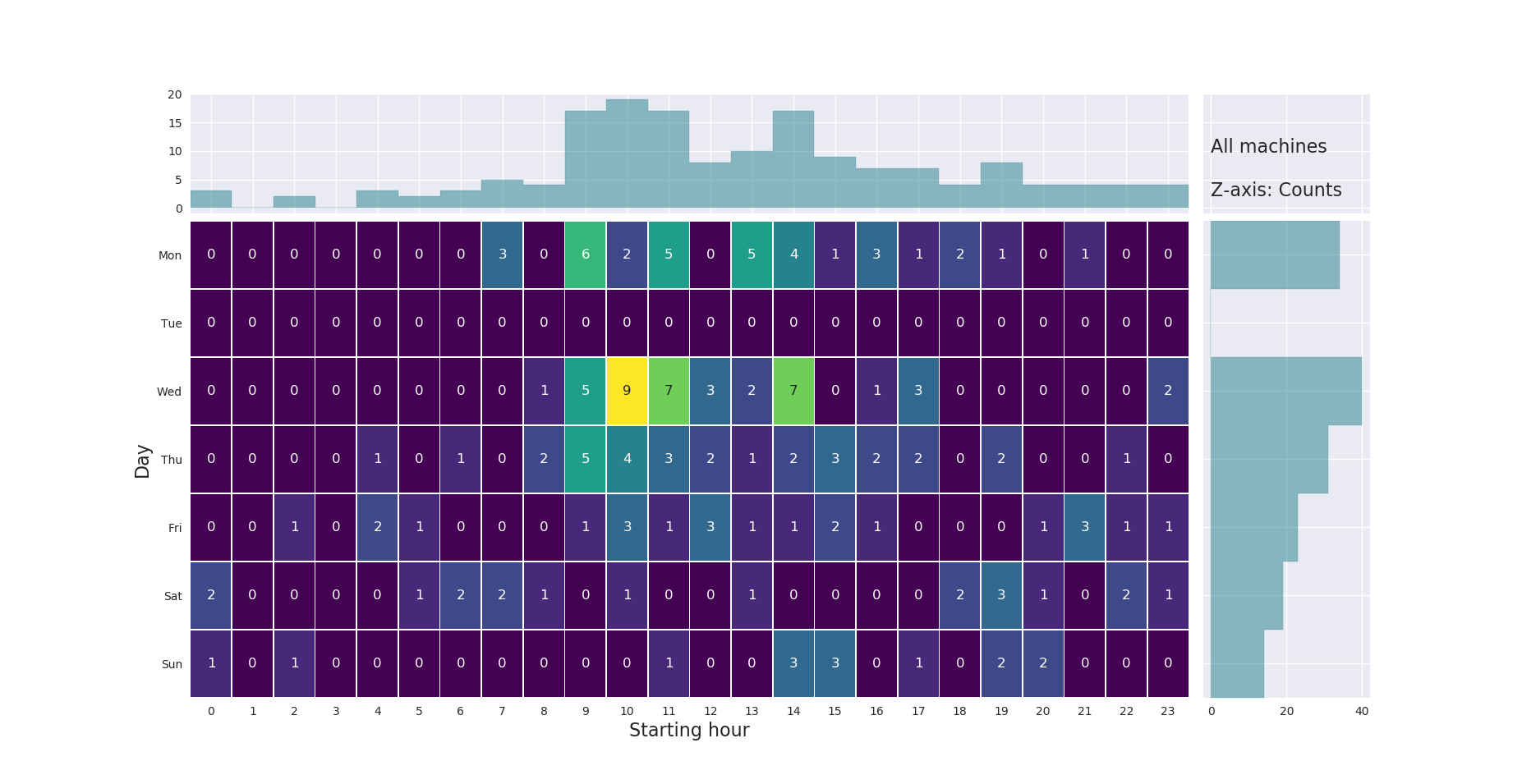} 
\captionsetup{width=0.68\linewidth}
\caption{Heatmap of downtime at MAXIV as a function of days and start hour for all three machines (I, R3, R1).  }
\label{fig:downtimeDayHour}
\end{figure*}

The database can be downloaded as a CSV file and analyzed in further detail. The heatmap in Figure~\ref{fig:downtimeCodeWeek} shows number of downtime events by code and week number for all three systems (I, R3, R1). The histograms on the margins are the projections of the heatmap onto the x and y-axis. The histogram on the right margin shows that certain event types are more frequent, and the upper histogram highlights how the downtime is distributed across the run period. 
The heatmap in Figure~\ref{fig:downtimeDayHour} shows the distribution of downtime events across days and hours (the hour during which the downtime started). The upper histogram shows that downtime is more likely to occur during daytime. This is likely because there are more users during this time, which leads to more parameters changing in the machines. During the spring run period of 2019, Tuesdays were designated as maintenance days, which entailed little or no delivery. The histogram on the right margin shows that Monday and Wednesday are the days with the most downtime events while the number of events decreases as the week progresses. We conclude from this that downtime is more frequent on Mondays when people resume their activities after the weekend and on Wednesdays when we have a complete restart of activities following the maintenance day. To minimize downtime, as well as to maximise the efficiency of machine studies periods,  the maintenance days have thus been moved to Mondays starting in the Autumn run of 2019.
\par
\begin{figure*}[!htb]
\includegraphics*[width=.59\textwidth]{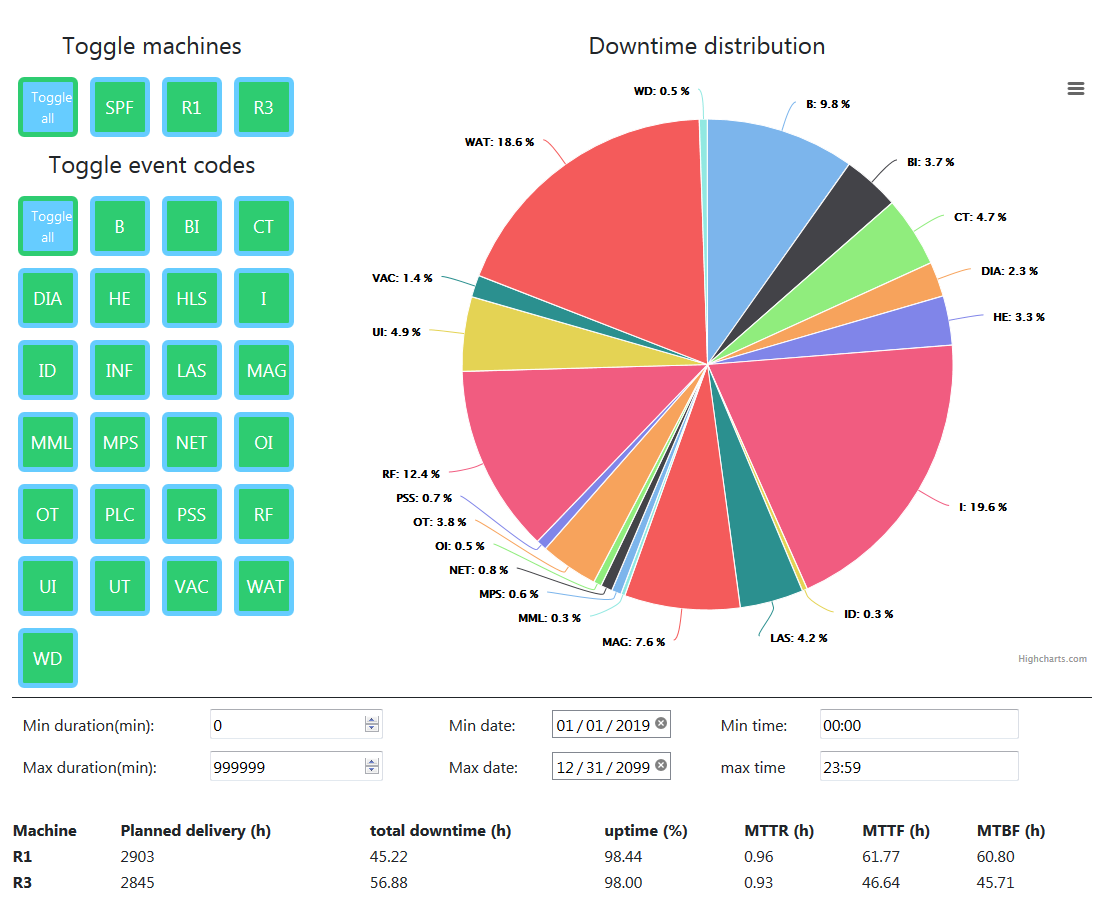} 
\captionsetup{width=.59\linewidth}
\caption{The front page of the downtime web-application.}
\label{fig:downtimeApp}
\end{figure*}

Figure~\ref{fig:downtimeApp} shows that a beam availability of 98.4\% was achieved for the 1.5~GeV storage ring (with 60h of MTBF) and of 98.0\% for the 3~GeV ring (with 40h of MTBF) during the first half of 2019. This can be compared to a beam availability of 96.7\% for the 1.5~GeV storage ring (with 59.6h of MTBF) and of 96.2\% for the 3~GeV ring (with 34.5h of MTBF) during 2018. The straightforwardness in which those numbers, as well as as the analyses plots can be obtained with the application have an important influencing factor for accelerator operations policies.

\section{CONCLUSION} %
\label{conclusion}
Several applications developed by the MAX IV accelerator operations group were presented. Most of the tools are for monitoring purposes, but the group has also developed software that directly improved the beam stability seen by the beam users. Specifically, the following software tools were presented: (i) a tool which displays the beam position's evolution over time
and the accelerator tunes' evolution, (ii) a tool that reduces the impact of changing orbit bumps for the beamlines, (iii) an application for adjusting the master oscillator's radio-frequency of a storage ring that is nearly transparent for users, and (iv) the downtime web-application used for registering and monitoring the availability and downtime of accelerators.
The development of these programs have significantly improved accelerator operations at the MAX IV laboratory.


\clearpage
\bibliography{bibliography.bib}

\end{document}